\documentclass[%
 reprint,
superscriptaddress,
showpacs,preprintnumbers,
 amsmath,amssymb,
 aps,
]{revtex4-1}

\usepackage{graphicx}
\usepackage{dcolumn}
\usepackage{bm}
\usepackage[colorlinks=true,linkcolor=blue,citecolor=blue,urlcolor=black]{hyperref}

\begin{document}

\title{Optical responses and optical activity of ultrathin films of topological insulator}

\author{Fariborz Parhizgar}\email{fariborz.parhizgar@ipm.ir}
\affiliation{School of Physics, Institute for Research in
Fundamental Sciences (IPM), Tehran 19395-5531, Iran}

\author{Ali G. Moghaddam}\email{agorbanz@iasbs.ac.ir}
\affiliation{Department of physics, Institute for advanced studies in basic science (IASBS), zanjan 45137-66731 Iran}
\affiliation{School of Physics, Institute for Research in
Fundamental Sciences (IPM), Tehran 19395-5531, Iran}

\author{Reza Asgari}
\affiliation{School of Physics, Institute for Research in
Fundamental Sciences (IPM), Tehran 19395-5531, Iran}
\date{\today}

\begin{abstract}
We investigate the optical properties of an ultrathin film of a topological insulator in the presence of an in-plane magnetic field. We show that due to the combination of the overlap between the surface states of the two layers and the magnetic field, the optical conductivity can show strong anisotropy. This leads to the effective optical activity of the ultrathin film by influencing the circularly polarized incident light. Intriguingly, for a range of magnetic fields, the reflected and transmitted lights exhibit elliptic character. Even for certain values almost linear polarizations are obtained, indicating that the thin film can act as a polaroid in reflection. All these features are discussed in the context of the time reversal symmetry breaking as one of key ingredients for the optical activity.
\end{abstract}
\maketitle

\section{Introduction}
Optical activity of a material which refers to its effect on the polarization of light passing through it in general originates from some symmetry breaking or anisotropy in the structure of materiels~\cite{Barron72, Barron81}. Magneto-optical effects, which are known for more than one and half century, have in particular their roots in the time reversal symmetry (TRS) breaking in the presence of a magnetic field.
Besides, magneto-optical effects have applications in measuring instruments, optical devices, and chemical characterizations \cite{Pershans67,Zak90}. Recently, investigations about the optical properties of low-dimensional and specially two dimensional (2D) systems are increased a lot motivated by synthesis of a variety of 2D materials including graphene \cite{Neto09}, monolayers of transition metal dichalcogenides \cite{Chhowalla}, and also the thin film of the topological insulators (TI) \cite{Hasan10,Qi11}. A great advantage of all these 2D systems is the tunability of their electronic structure by electric or magnetic fields which leads to many promising applications in nanoelctronics, spintronics as well as optical and photonic devices.
\par
Among all of above mentioned new materials, the discovery of the TI is of more fundamental importance since it opens new lines of investigation in the theory of condensed matter physics~\cite{Fu07,Qi08}. In particular by reformulating the band theory, providing better understanding of topological orders, and contributing to the field of the quantum information and quantum computing, topological insulators are in the heart of interests in recent years \cite{Hasan10,Qi11}. Topological insulators in three dimension are insulating materials in which the strong spin orbit interaction causes the band inversion~\cite{Tibook}. At the surface of the TI, the bands turn back to their natural order and subsequently topologically protected surface states emerge. These states usually governed by some relativistic dynamics persist a robust gapless dispersion unless the TRS is broken by applying magnetic fields or in the presence of magnetic impurities.
\par
The strong sensitivity of the TI to any perturbation which breaks the TRS leads to interesting phenomena in their magneto-optical responses \cite{Suchkov10,Jenkins10,Aguilar12}.
A key finding has been provided by Tse and MacDonald revealing a giant magneto-optical Kerr effect besides a universal
Faraday angle, $\theta_{\text F}=\arctan(\alpha)$ where the fine structure constant is $\alpha=e^2/\hbar c$, at the long wavelength limit~\cite{Tse10,Tse11,Hasan}. They considered a thin film of the TI in the presence of a perpendicular magnetic field which can be described by the microscopic 2D massless Dirac model. Then using the linear response theory rather than topological field theory method \cite{Qi08}, the optical conductivity and the magneto-optical responses have been obtained. As a result of the interplay between confinement and the Hall effect of Dirac modes, only quantized currents are induced by the incident electromagnetic waves and subsequently the reflected and transmitted waves show an unexpected behavior. However, the TRS breaking and the gap opening due to the magnetic field play an important role in the observation of these optical phenomena.
\par
Remarkably, the optical conductivity of the TI surface states and a thin-film of the TI have been studied theoretically~\cite{theory, Rostami14, peres13} and the universal value for the the optical conductivity, $\sigma_0=e^2/4\hbar$ has been obtained. Moreover, the optical properties of bismuth-based topological insulators has been experimentally explored~\cite{exp}.
\par
In this paper we analyze the optical responses of ultra-thin films of the TI in the presence of the in-plane magnetic field, $B$. Due to the overlap between the surface states, a gap $\Delta$ opens in the dispersion relation even without the TRS breaking.
The effect of the in-plane magnetic field is a constant difference in the vector potential of the surface states at the two sides. Without hybridization of surface states such difference can be gauged out without any physical effect, however the overlap between them prohibits independent gauge transformations for two sides. Therefore, the combination of the in-plane magnetic field and hybridization leads to the interesting behavior of their optical responses.

We show that the optical conductivity tensor exhibits a profound anisotropy and its longitudinal components $\sigma_{xx}$ and $\sigma_{yy}$ change drastically by the variation of the imposed field. Competition between the hybridization and an energy scale proportional to $B$ leads to two different phases of the system. In small $B$ values, the thin film exposes a gapped dispersion relation, however at some certain value of the magnetic field, the gap closes and the system enters into a gapless phase. Some qualitative differences in the behavior of the optical conductivity are found for those two phases.
Then we investigate the properties of the reflected and transmitted electromagnetic waves when a right-handed circularly polarized light is incident normally on the system. Intriguingly, we obtain, for some certain range of the $B$ around the band structure phase transition, the reflected light becomes elliptic and even nearly linearly polarized while the transmitted one always remains close to the circular polarization. Far away from the transition \emph{i.e.} for very small $B$ or $\Delta$, the time reversal symmetry restores and the system behaves similar to a simple isotropic metal.
\par
This paper is organized as follows. Sec. \ref{model} is devoted to the theoretical model and basic formalisms. First, a model Hamiltonian of the system is introduced and then the conductivities are calculated by using Kubo formalism. The model for the optics and propagation of the electromagnetic waves through thin film is given. In Sec. \ref{result} we present our numerical results.
Finally, we conclude our main results in Sec. \ref{conc}.

\section{Model and Theory}
\label{model}
Ultrathin films of the TI have two surfaces with opposite helical states which can be hybridized by a tunneling parameter $\Delta$. The Hamiltonian of such a thin film can be written as \cite{Zyuzin,Sergey},
\begin{equation}
H=\begin{pmatrix} h({\bf k}) & \Delta \\ \Delta & -h({\bf k})
\end{pmatrix},
\end{equation}
where $h({\bf k})= \hbar v\hat{z}(\hat{{\bf \sigma}}\times{\bf k})$ is the Dirac Hamiltonian describing the surface states of the TI. Here $\hat{z}$ shows the direction perpendicular to the upper surface, $\hat{{\bf \sigma}}$ indicates the Pauli matrices for the real spin and $v=5\times 10^5 m/s$ is the Fermi velocity of the helical states in the TI. The hybridization parameter, $\Delta$ is allowed for the helical states of the same kind and depends on the thickness of the sample. It has been experimentally shown that $\Delta$ varies from $0.01$ to $0.25 eV$ for thicknesses of $d=10$ nm to ultrathin $2$ nm films~\cite{Zhang}. The new Hamiltonian describes massive Dirac Fermions with the gapped dispersion  $\varepsilon(k)=\pm\sqrt{\hbar^2 v^2k^2+\Delta^2}$ which is in contrast with massless Dirac states present at a single surface of a thick TI.
\par
\begin{figure}
\includegraphics[width=1.0\linewidth]{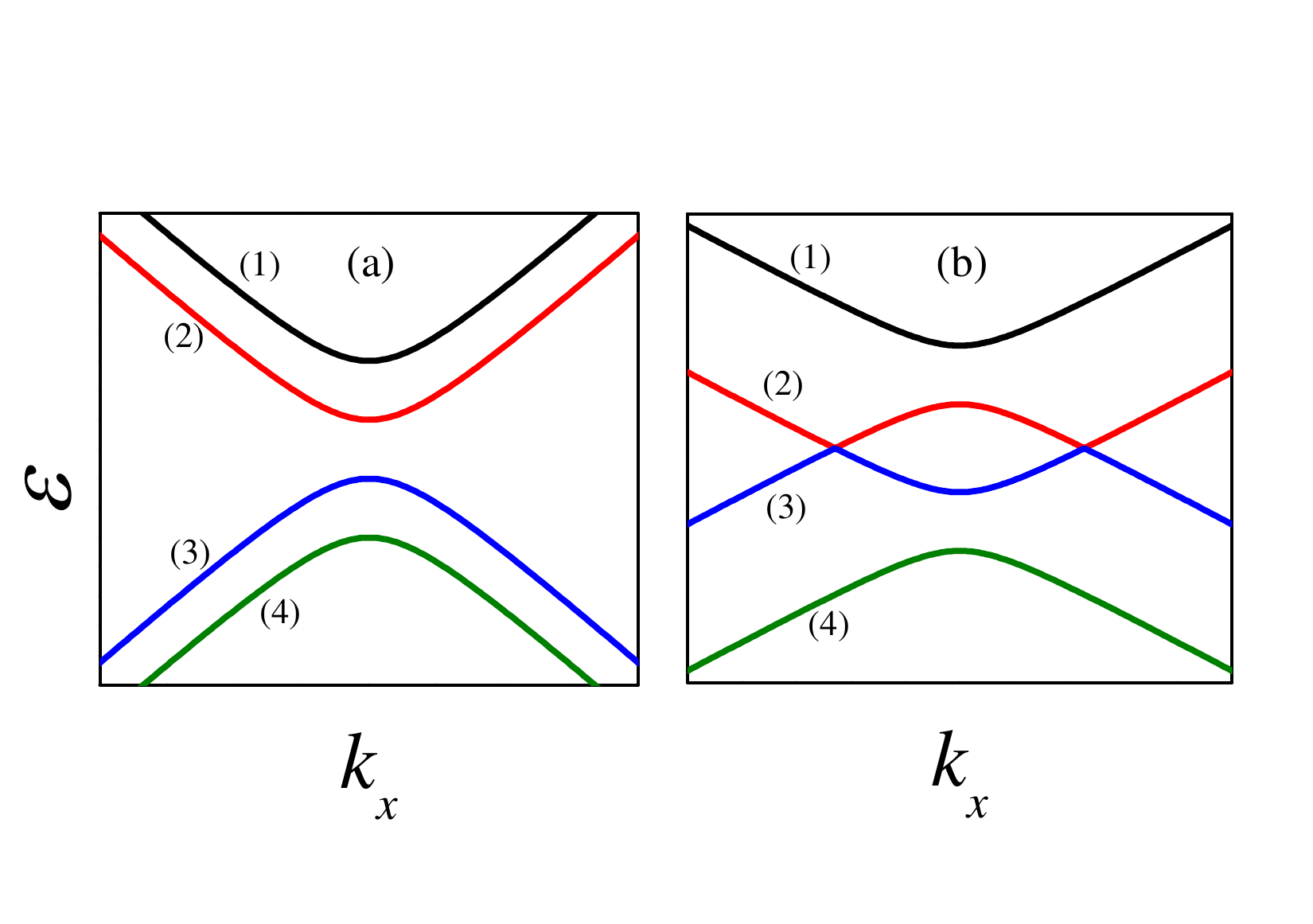}
\caption{(Color online) Dispersion of the topological insulator thin film along in the $k_x$ direction at the presence of an in-plane magnetic field and a hybridization $\Delta$.
Left and right panels correspond to (a) $\varepsilon_B<\Delta$ and (b) $\varepsilon_B>\Delta$, respectively.
\label{fig:schem}}
\end{figure}
Applying a magnetic field parallel to the single surface of the TI just shifts the position of the Dirac cone in the momentum space and can be gauged out without any physical implications. While in the case of ultrathin films, as mentioned above, the in-plane magnetic field induces two different gauge shifts in two different surfaces and subsequently the band dispersion can be affected by $B$. For the sake of definiteness, we consider ${\bf B}=B_y \hat{y}$ which leads to the gauge field $A=zB_y\hat{x}$ and as a result one can write the new Hamiltonian as~\cite{comment},
\begin{equation}\label{eq:Hamil}
H=\begin{pmatrix} h({\bf k}-{\bf q}_B) & \Delta \\
\Delta & h({\bf k}+{\bf q}_B)
\end{pmatrix},
\end{equation}
where ${\bf q}_B=eB_yd/2~ \hat{x}$ and the layers is separated by distance $d$. The energy dispersion has the form $\varepsilon_{s,t,{\bf k}}=s\sqrt{\hbar^2 v^2k^2\sin^2 \phi+g_t^2(k)}$ with $g_t(k)=\hbar vq_B+t\sqrt{\hbar^2 v^2k^2\cos^2\phi+\Delta^2}$. Here $s=\pm 1$ shows the valance $(s=-1)$ and conduction $(s=+1)$ bands and $t=\pm 1$ indicates the bands with larger and smaller gaps, respectively. The presence of the magnetic field causes to the anisotropy of the dispersion relation indicated by the explicit dependence on the momentum direction angle $\phi=\arctan(k_y/k_x)$, while for $B=0$ the band structure becomes completely isotropic, as it would be expected.
The dispersion relation along the $x$ direction exhibits two different phases according to the value of the $q_B$, equally discussed in Ref.~[\onlinecite{Zyuzin}]. This is due to the competition between the gap size $\Delta$ and the magnetic cyclotron energy $\varepsilon_B=\hbar vq_B$ as shown in Fig.~\ref{fig:schem}. In small magnetic fields where $\varepsilon_B<\Delta$, the system is gapped while at $\varepsilon_B>\Delta$, two subbands intersect subsequently each other at the zero energy and close the gap. After straightforward algebra, the four eigenfunctions of the Hamiltonian Eq.~\eqref{eq:Hamil} indicated by $|s,t \rangle$ are obtained as
\begin{eqnarray}\label{eq:wavef}
|s,+ \rangle=\frac{1}{\sqrt{2}}\begin{pmatrix}
i \sin \frac{\gamma}{2} e^{i\theta_+}\\
-s  \sin \frac{\gamma}{2} \\
i s \cos \frac{\gamma}{2} \\
-\cos \frac{\gamma}{2} e^{i\theta_+}
\end{pmatrix} \,,
|s,- \rangle=\frac{1}{\sqrt{2}}\begin{pmatrix}
i \cos \frac{\gamma}{2} e^{i\theta_-}\\
-s  \cos \frac{\gamma}{2} \\
-i s \sin \frac{\gamma}{2} \\
\sin \frac{\gamma}{2} e^{i\theta_-}
\end{pmatrix},\nonumber\\
\end{eqnarray}
with the newly defined parameters $\gamma$ and $\theta_t$ in order that  $\sin\gamma=\Delta/\sqrt{\hbar^2 v^2 k^2\cos^2\phi+\Delta^2}$, $\sin\theta_t=\hbar v k\sin\phi/\sqrt{\hbar^2 v^2 k^2\sin^2\phi+g_t^2(k)}$.

\subsection{Optical conductivity of the thin film TI}
In order to calculate the optical conductivity tensor $\sigma_{\alpha\beta}(\omega)$, we invoke the linear response theory and the Kubu formula given by~\cite{Sinitsyn,Tse11},
\begin{equation}\label{eq:Kubo}
\sigma_{\alpha\beta}(\omega)=i\sum_k\sum_{\mu \mu' }\frac{f_{k,\mu}-f_{k,\mu'}}{\varepsilon_{k,\mu}-\varepsilon_{k,\mu'}}\frac{{\cal J}^{\mu \mu'}_{\alpha\beta}(k)}{\hbar \omega+\varepsilon_{k\mu}-\varepsilon_{k\mu'}+i/2\tau_s}
\end{equation}
where $\mu,\mu'$ denote the different bands and $\alpha,\beta$ indicate $x,y$ axes. The Fermi-Dirac distribution function is indicated by $f_{k,\mu}$ at the Fermi level $\mu$ and $\tau_s=1/\Gamma$ is a tiny residual scattering time due to impurities. The tensor quantity
${\cal J}^{\mu\mu'}_{\alpha\beta}(k)=\langle k,\mu|j_\alpha|k,\mu' \rangle \langle k,\mu'|j_\beta|k,\mu \rangle$ is given by the components of the current operator $j_{\alpha}=e\partial H / \partial k_{\alpha}$ which for the ultrathin film of the TI reads as,
\begin{equation}
j_{x(y)}=\begin{pmatrix} \mp\sigma_{y(x)} &0 \\ 0& \pm \sigma_{y(x)}
\end{pmatrix}.
\end{equation}
In our model, since there is no longer perpendicular magnetic field or magnetization, the Hall conductivity vanishes, $\sigma_{xy}=0$ however, the in-plane field makes the conductivity strongly anisotropic namely $\sigma_{xx}\neq \sigma_{yy}$. We include only the interband transitions at zero-Fermi energy and the contribution of the intraband transitions, which leads to a Drude-like term, is no longer relevant in this study since the momentum relaxation time is assumed to be very large. This approximation is valid at low-temperature and a quite clean sample. We also do not consider the bound state of exciton in the system.

Using the wave functions by Eq.~\eqref{eq:wavef} and the basis as $\psi_{s,t}^\dagger=(\psi_{+,+}^\dagger,\psi_{+,-}^\dagger,\psi_{-,-}^\dagger,\psi_{-,+}^\dagger)$, the longitudinal parts of the quantity ${\cal J}^{\mu\mu'}_{\alpha\alpha}(k)$ are given by,
\begin{widetext}
\begin{equation}\label{eq:J_xx}
{\cal J}_{xx}(k)=\begin{pmatrix}
\cos^2\gamma\cos^2 \theta_+  & \sin^2 \gamma \cos^2 \frac{\theta_++\theta_-}{2}  & \sin^2\gamma\sin^2 \frac{\theta_++\theta_-}{2}  & \cos^2\gamma\sin^2 \theta_+  \\
\sin^2\gamma\cos^2 \frac{\theta_++\theta_-}{2}  & \cos^2\gamma\cos^2 \theta_-  &
\cos^2\gamma\sin^2 \theta_-  &\sin^2\gamma\sin^2 \frac{\theta_++\theta_-}{2}   \\
\sin^2\gamma\sin^2 \frac{\theta_++\theta_-}{2}  & \cos^2\gamma\sin^2 \theta_-  &
\cos^2\gamma\cos^2\theta_- &\sin^2\gamma\cos^2\frac{\theta_++\theta_-}{2}  \\
\cos^2\gamma\sin^2\theta_+ & \sin^2\gamma\sin^2\frac{\theta_++\theta_-}{2} & \sin^2\gamma\cos^2\frac{\theta_++\theta_-}{2} & \cos^2\gamma\cos^2\theta_+
\end{pmatrix}
\end{equation}
\end{widetext}
and
\begin{eqnarray}\label{eq:J_yy}
{\cal J}_{yy}(k)=
\begin{pmatrix}
\sin^2\theta_+&0&0&\cos^2\theta_+\\
0&\sin^2\theta_-&\cos^2\theta_-&0\\
0&\cos^2\theta_-&\sin^2\theta_-&0\\
\cos^2\theta_+&0&0&\sin^2\theta_+
\end{pmatrix}.
\end{eqnarray}
Based on the Eqs. \eqref{eq:J_xx} and \eqref{eq:J_yy}, we are able to explore the selection rules of the optical conductivity in the system. For the sake of clarification, let us label the eigenfunctions by
$(s,t)=(+,+),(+,-),(-,-),(-,+)$ and corresponding bands shown in Fig. \ref{fig:schem} with numbers from $1$ to $4$, respectively. As seen in Eq.~\eqref{eq:J_yy}, the transition $3 \rightarrow 1$ and $4 \rightarrow 2$ are forbidden. Also later on, we show numerically that although all transitions are allowed for $\sigma_{xx}$ but the transition $3 \rightarrow 2$ and $4 \rightarrow 1$ are negligible and has a minor effect.
Note that in this paper, we assume $E_{\text F}=0$ so the transitions $3\rightarrow 4$ and $1\rightarrow 2$ are unimportant regardless of the ${\cal J}$ values.

The phenomenological scattering rate $\Gamma$ is chosen to be $\Gamma=1$ meV which originates from possible impurity and defects in the system~ \cite{Tse11}. Remarkably, we note that in the undoped case, $E_{\text F}=0$ the Kubo formula only includes the interband conductivity and thus the intraband part ($\mu=\mu'$) vanishes.

\subsection{Propagation of electromagnetic waves through the thin film TI}
Here, we explore the properties of the reflected and transmitted electromagnetic waves of an incident wave
which are completely given by the optical conductivity tensor. We assume that the thin film lies in the $x-y$ plane and the incident and scattered lights which are plane waves propagate in the $z$-direction normal to the thin film plane. The intensity, phase and polarization of the electromagnetic waves can be obtained from their electric fields ${\bf E}^{i}$, ${\bf E}^{r}$, and ${\bf E}^{t}$ at $z=0$ corresponding to the incident, reflected, and transmitted parts. We should remind that the electric field here is a complex two-component vector ${\bf E}=E_x\hat{\bf x}+E_y\hat{\bf y}$. The reflected and transmitted waves are related to the incident one with linear relations ${\bf E}^{r}=\hat{\bf r} {\bf E}^{i}$ and ${\bf E}^{t}=\hat{\bf t}{\bf E}^{i}$ in which $\hat{\bf r}$ and $\hat{\bf t}$ are $2\times 2$ complex matrices in the two dimensional space corresponding to the $x-y$ plane. If we ignore the thickness of the thin film in comparison with the wavelength and denote the field in the two sides of the film with ${\bf E}^{(u)}$ and ${\bf E}^{(d)}$, then the boundary conditions for the electric and magnetic fields will be ${\bf E}^{(u)}={\bf E}^{(d)}$ and $-i\hat{\tau}_y({\bf H}^{(u)}-{\bf H}^{(d)})=(4\pi/c){\bf J}=(4\pi/c)\hat{\bm \sigma} {\bf E}$. Assuming the incident light hits the upper side of the film, the electric fields of the two sides are given by ${\bf E}^{(u)}={\bf E}^{i}+{\bf E}^{r}$ and ${\bf E}^{(d)}={\bf E}^{t}$ and magnetic fields follows the corresponding relations. Adding to these relations and the boundary conditions, the fact that for the vacuum at each side ${\bf H}=\hat{\bf z}\times {\bf E}$, we derive the transmission and reflection matrices as below,
\begin{equation}\label{eq:r,t}
\hat{\bf r}=\begin{pmatrix}
\frac{-\tilde{\sigma}_{xx}}{\tilde{\sigma}_{xx}+2}&0\\0&\frac{-\tilde{\sigma}_{yy}}{\tilde{\sigma}_{yy}+2}
\end{pmatrix} ~,~ \hat{\bf t}=\begin{pmatrix}
\frac{2}{\tilde{\sigma}_{xx}+2}&0\\0&\frac{2}{\tilde{\sigma}_{yy}+2}
\end{pmatrix}
\end{equation}
where $\tilde{\sigma}_{\alpha\beta}=(4\pi/c)\sigma_{\alpha\beta}$ denotes the dimensionless conductivity.
\par
Now, we suppose a right-circularly polarized electromagnetic wave of the form ${\bf E}^{i}=E_0(\hat{\bf x}+i\hat{\bf y})/\sqrt{2}$ is incoming to the TI plane. Due to the anisotropic conductivity ($\sigma_{xx}\neq \sigma_{yy}$) which is a consequence of the in-plane magnetic field, the reflected and transmitted waves will be of elliptical polarization as,
\begin{equation}
\label{eq:ref-tr}
\begin{array}{l}
{\bf E}^{r}=E_0 (r_{xx}\hat{\bf x}+ir_{yy}\hat{\bf y})/\sqrt{2}\\
{\bf E}^{t}=E_0 (t_{xx}\hat{\bf x}+it_{yy}\hat{\bf y})/\sqrt{2}
\end{array}
\end{equation}
in which the components of matrices $\hat{\bf r}$ and $\hat{\bf t}$ are defined in Eq. \eqref{eq:r,t}.
It is more convenient to use the circular polarization bases given by $\hat{\bm \epsilon}_{\pm}=(\hat{\bf x}\pm i\hat{\bf y})/2$ indicating the right- and left-handed circular polarizations, respectively \cite{Jackson}. The general elliptically polarized wave is thus described with ${\bf E}=E_{+}\hat{\bm \epsilon}_+ +E_{-}\hat{\bm \epsilon}_-$.

The eccentricity and the rotation angle of the polarization ellipse are given by $e_{P}=2/(\sqrt{|E_{+}/E_{-}|}+\sqrt{|E_{-}/E_{+}|})$ and $\alpha_{P}=(\varphi_{+}-\varphi_{-})/2$ in which $\varphi_{\pm}=\arg E_{\pm}=\arctan({\rm Im}E_{\pm}/{\rm Re}E_{\pm})$.
Then by invoking Eq. \eqref{eq:ref-tr}, the eccentricity and rotation angle of the reflected electromagnetic wave are,
\begin{eqnarray}
e_{P}^{r}&=&\frac{2}{\sqrt{|\frac{r_{xx}-r_{yy}}{r_{xx}+r_{yy}}|}+\sqrt{|\frac{r_{xx}+r_{yy}}{r_{xx}-r_{yy}}|}},
\label{eq:eccent} \\
\alpha_{P}^{r}&=&\frac{1}{2} \arg(\frac{r_{xx}+r_{yy}}{r_{xx}-r_{yy}}),
\label{eq:angle}
\end{eqnarray}
and the transmitted part obeys the corresponding relation with $r_{\alpha\alpha}$ replaced by $t_{\alpha\alpha}$.
It is clear that the anisotropy in the optical conductivity leads to the difference in the reflection and transmission components ($r_{\alpha\alpha}$ and $t_{\alpha\alpha}$). In general, we have elliptically polarized scattered waves with a non-vanishing eccentricity. We should remind that the eccentricity is a measure of ellipticity, varying from $0$ for circular polarizations to $1$ corresponding to the linearly polarized waves.
\par
Using the matrices $\hat{\bf r}$ and $\hat{\bf t}$, we obtain the intensities of reflected and transmitted waves as,
\begin{equation}
\begin{array}{l}
\frac{I_r}{E_0^2}=\frac{1}{2}\sum_\alpha \frac{
|\tilde{\sigma}_{\alpha\alpha}|^2
}{(2+{\rm Re}\tilde{\sigma}_{\alpha\alpha})^2+({\rm Im}\tilde{\sigma}_{\alpha\alpha})^2}\\
\frac{I_t}{E_0^2}=\frac{1}{2}\sum_\alpha \frac{4}{(2+{\rm Re}\tilde{\sigma}_{\alpha\alpha})^2+({\rm Im}\tilde{\sigma}_{\alpha\alpha})^2}
\end{array}
\end{equation}
As a result, the absorbed wave intensity is given by,
\begin{equation}
\label{absorb}
\frac{I_{\rm abs}}{E_0^2}=\sum_\alpha \frac{
{2\rm Re}\tilde{\sigma}_{\alpha\alpha}
}{(2+{\rm Re}\tilde{\sigma}_{\alpha\alpha})^2+({\rm Im}\tilde{\sigma}_{\alpha\alpha})^2}
\end{equation}
which is always positive due to the fact that $Re{\sigma_{\alpha\alpha}}>0$. This itself originates from the second law of thermodynamics revealing as power dissipation instead of the power generation.

\section{Numerical Result and discussion}
\label{result}
In this section, main results of our study are presented. First, the results of calculations for the optical conductivity of the ultrathin film of the TI in the presence of the in-plane magnetic field are shown as functions of the frequency $\omega$ and also the magnetic energy scale $\hbar v q_B$. In order to see the optical activity of the TI, we assume a circular electromagnetic wave applied perpendicularly to the sample's plane. The properties of the reflected and transmitted parts are uncovered by studying their polarizations and in particular the degree of the ellipticity and the rotation angle of the polarization ellipse. Finally, the absorption at different values of the magnetic fields is discussed. In numerical calculation we set $\Delta = 50$ meV and $\Gamma=1$ meV.

\subsection{Optical conductivity}
As we discussed in the previous section, the low-energy band structure of the system consists of four different bands which is gapped in the case that $\varepsilon_B=\hbar v q_B<\Delta$ while the gap vanishes in the case that $\varepsilon_B\geq\Delta$. We consider the undoped system at zero temperature so that, there are only four different types of transitions between those bands. The first and second types of transitions occur between the bands $4$ to $1$ and $3$ to $2$, respectively which have the largest and smallest gaps.
The two other transitions are between $3$ to $1$ and $4$ to $2$ having similar energy differences. Since at zero temperature and undoped situation, the two bands $3$ and $4$ are filled and the others are completely empty, the two remaining transitions are forbidden.

\begin{figure}
\includegraphics[width=1.0\linewidth]{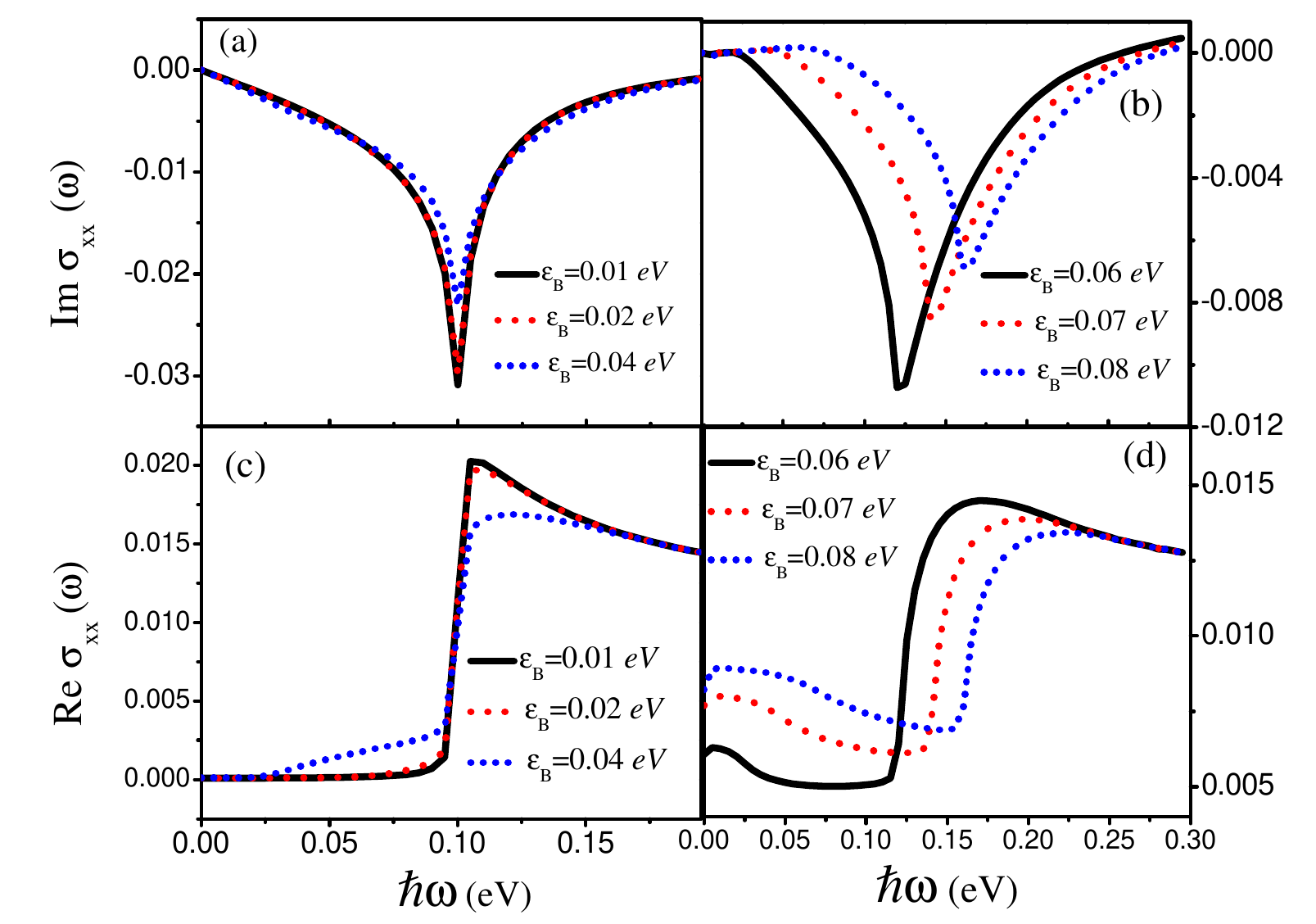}
\caption{(Color online) Imaginary (a,b) and real (c,d) parts of the optical conductivity component $\sigma_{xx}$, scaled by $4\pi/c$ as a function of the energy when $\varepsilon_B<\Delta$ (a,c) and $ \varepsilon_B>\Delta$ (b,d). In (a,c) the real part of the $\sigma_{xx}$ is almost zero for frequencies below $2\Delta$, and the imaginary part shows a peak at $\omega=2\Delta$ and the position of peaks and steplike configuration
do not change due to the $\varepsilon_B$. However, the peak position does move to higher energy by increasing $\varepsilon$ in the case (b,d) and ${\rm Re}\sigma_{xx}$ shows a small peak at very low frequencies $\omega$ according to the transition between $2$ and $3$ band which was negligible for $\varepsilon_B<\Delta$.
\label{fig:xx}}
\end{figure}

Figure \ref{fig:xx} shows the imaginary and real parts of optical conductivity component $\sigma_{xx}$ (scaled by $4\pi/c$) of the ultrathin film of the TI in the presence of the in-plane magnetic field in the $y$-direction.
In the gapped phase, the real part of the $\sigma_{xx}$ is almost zero for frequencies below $2\Delta$, but for higher frequencies the dissipation channel corresponding to the transitions
from $3$ to $1$ and from $4$ to $2$ takes the role. On the same ground, the imaginary part shows a significant peak at $\omega=2\Delta$ where the most profound photon-induced scattering of electrons between the mentioned bands are occurred. The value of the peak position for both imaginary and real parts of the longitudinal coefficient are equal as they are related by the Kramers-Kroning relations and in addition, the position of peaks and steplike configuration
do not change due to the $\varepsilon_B$ where $\varepsilon_B<\Delta$. However, the peak position does move to higher energies by increasing $\varepsilon_B$ in the case that $\varepsilon_B>\Delta$. Therefore, the position of peaks or steplike configuration of the dynamical conductivity, can be controlled by the in-plane magnetic field.
Our results in Fig. \ref{fig:xx}(a),(c) indicate that although other transitions are not prohibited according to Eq. \eqref{eq:J_xx}, they have negligible contribution to this component of optical conductivity.
When we pass through the transition point $\varepsilon_B>\Delta$ and the gapless phase is reached, the behavior of the conductivity changes as shown in Fig. \ref{fig:xx}(b),(d).
Here ${\rm Re}\sigma_{xx}$ shows a small peak at very low frequencies $\omega$ according to the transition between $2$ and $3$ band which was negligible for $\varepsilon_B<\Delta$. Another more profound jump appears due to the previously mentioned transitions from $3$ to $1$ and from $4$ to $2$ which the energy differences are not $2\Delta$ anymore and increases as $2\varepsilon_B$ by the variation of $B$. The imaginary part ${\rm Im}\sigma_{xx}$, subsequently, shows strong peak at $\omega=2\varepsilon_B$.

\begin{figure}
\includegraphics[width=1.0\linewidth]{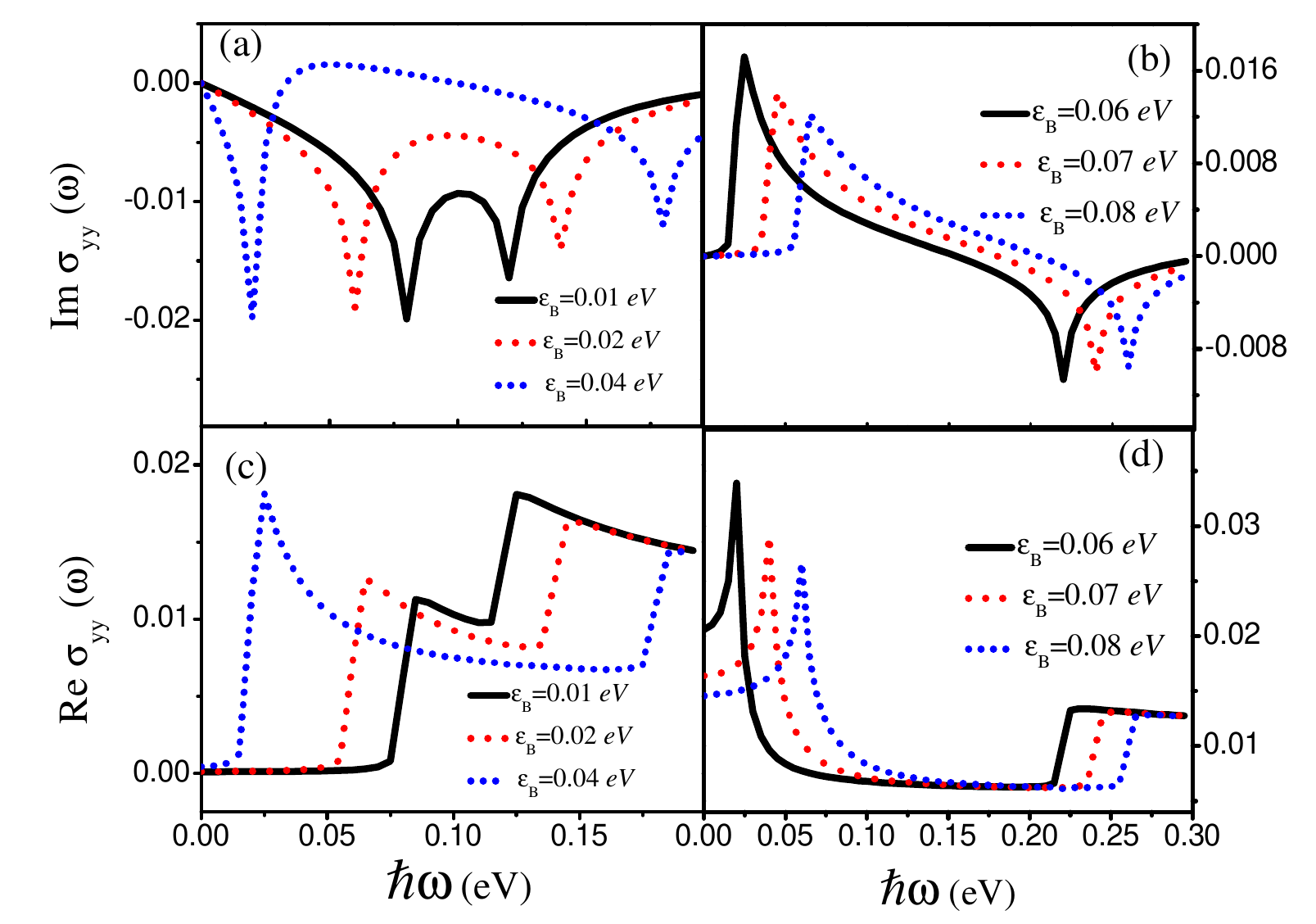}
\caption{(Color online) Imaginary (a,b) and real (c,d) parts of the optical conductivity component $\sigma_{xx}$, scaled by $4\pi/c$ as a function of the enegy when $\varepsilon_B<2\Delta$ (a,c) and $\varepsilon_B>2\Delta$ (b,d). Notice that two peaks (jumps) are always present in ${\rm Im}\sigma_{xx}$ (${\rm Re}\sigma_{xx}$) corresponding to the smallest and largest energy differences as $2(\Delta+\varepsilon_B)$ and $2|\Delta-\varepsilon_B|$.
The real part of the conductivity starts from a finite value and it decreases with the magnetic field.
\label{fig:yy}}
\end{figure}

When one explores the conductivity component $\sigma_{yy}$ the selection rules change. According to Eq. \eqref{eq:J_yy}, there is no transition from $3$ to $1$ and $4$ to $2$. Other two transitions at zero doping occur at two different frequencies. Figure \ref{fig:yy} shows the behavior of $\sigma_{yy}$ (scaled by $4\pi/c$) at two different regimes of $\varepsilon_B<\Delta$ and $\varepsilon_B>\Delta$ as a function of $\omega$. First for $\varepsilon_B<\Delta$ (panels $a,c$) two peaks (jumps) are always present in ${\rm Im}\sigma_{xx}$ (${\rm Re}\sigma_{xx}$) corresponding to the smallest and largest energy differences: $2(\Delta+\varepsilon_B)$ and $2|\Delta-\varepsilon_B|$.
So by increasing the magnetic field the two peaks (jumps) become closer and reach each other at the transition point $\varepsilon_B=\Delta$. In the $\varepsilon_B>\Delta$ region as shown in Fig. \ref{fig:yy}(b),(d) the first peak disappears owing to the closure of the smallest gap. The real part of the conductivity starts from a finite value where this value decreases with the magnetic field. In addition, the behavior of imaginary part of the conductivity is different and it is mostly positive in contrast with the gapped phase.

\begin{figure}
\includegraphics[width=1.0\linewidth]{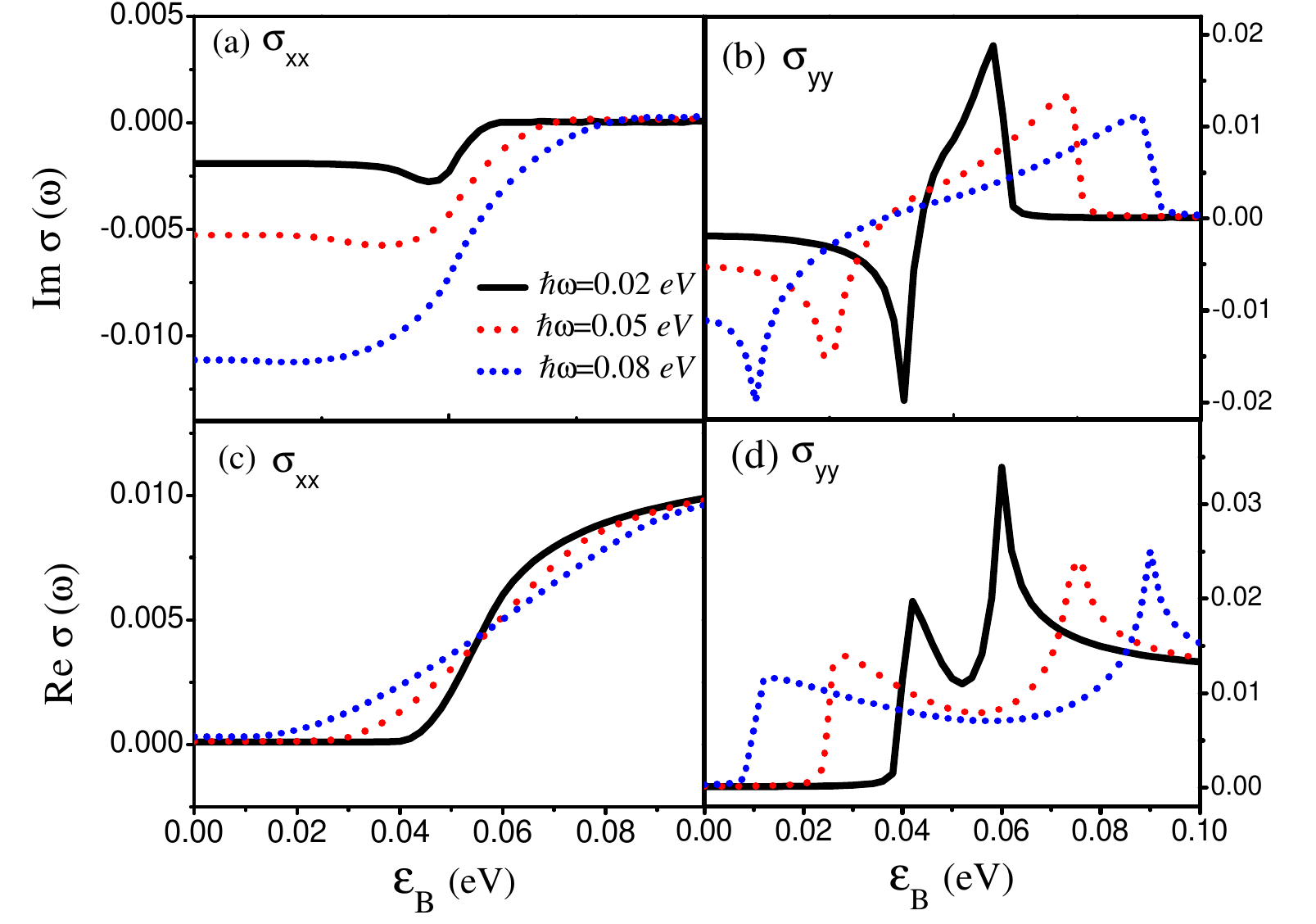}
\caption{(Color online) Imaginary and real parts of the optical conductivity, scaled by $4\pi/c$, of the thin film topological insulator as a function of magnetic field energy, $\varepsilon_B$.
\label{fig:q}}
\end{figure}
\par
In order to conclude these results we also plot the conductivities $\sigma_{xx}$ and $\sigma_{yy}$ as a function of $\varepsilon_B$ for different values of $\omega$ in Fig. \ref{fig:q}. As seen in Fig. \ref{fig:q}(a) around the transition point $\varepsilon_B=\Delta$ the behavior of the imaginary part of the conductivity $\sigma_{xx}$ changes almost abruptly from negative values to near zero ones.
As one may expect the dissipative component ${\rm Re}\sigma_{xx}$ is much larger in the gapless phase ($\varepsilon_B>\Delta$) in comparison to the gapped one ($\varepsilon_B<\Delta$), a behavior which is indicated in Fig. \ref{fig:q}(c). Finally the variation of imaginary and real parts of $\sigma_{yy}$ with magnetic field are shown in Fig. \ref{fig:q}(b),(d). Both parts reveal two peaks which one of them occurs
at $\varepsilon_B<\Delta$ and the other at larger magnetic fields $\varepsilon_B>\Delta$. Moreover although the real part
usually takes higher values for gapless case $\varepsilon_B>\Delta$ similar to $\sigma_{xx}$, however, ${\rm Im} \sigma_{yy}$ changes sign close to the transition point showing almost antisymmetric form around $\varepsilon_B=\Delta$.
\par
As a final remark at this part we note that the variations and also anisotropic in the conductivity tensor take place at the vicinity of transition points where both the magnetic field energy $\varepsilon_B$ and the overlap $\Delta$ are finite. When either $\varepsilon_B$ or $\Delta$ becomes much smaller than the other, the conductivities reaches some limiting values which are the same for $\sigma_{xx}$ and $\sigma_{yy}$. In particular the real parts of both components vanish at very small fields $\varepsilon_B\ll\Delta$ and the imaginary parts reache zero for small overlaps $\Delta\ll\varepsilon_B$. These are clear signatures of the fact that for vanishing $\varepsilon_B$ or $\Delta$ in fact the time reversal symmetry is not broken since either there is no magnetic field or it has no physical effect and can be omitted by a gauge transformation.

\subsection{Optical activity of ultathin films of the TI}
\begin{figure}[t]
\includegraphics[width=1.0\linewidth]{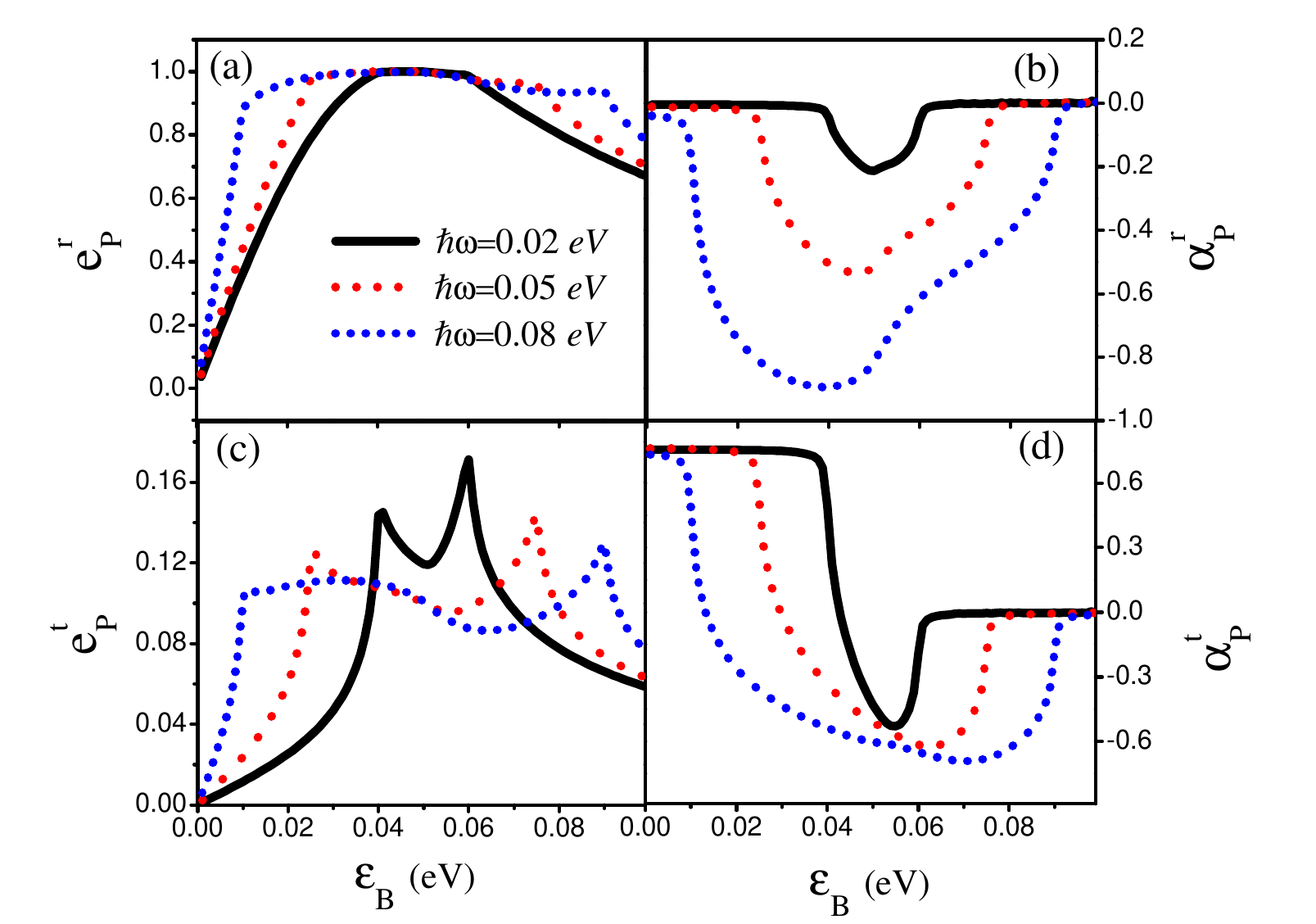}
\caption{(Color online) Eccentricity and phase shift (inset) of the reflected electromagnetic wave as a function of the magnetic field potential $\varepsilon_B$. The legend in (b,c,d) are the same as (a).
\label{fig:ref}}
\end{figure}

In the following, we would like to investigate the effect of the magnetic field on the optical activity of the ultrathin films when a circularly polarized light hits the film in a normal direction.
This is revealed by the phase shift $\alpha_{P}$ and eccentricity $e_{P}$ of the reflected and transmitted electromagnetic waves. Figure \ref{fig:ref} shows the behavior of the parameters $e_P$ and $\alpha_P$ defined in Eqs. \eqref{eq:eccent} and \eqref{eq:angle} for the reflected and transmitted components, respectively. As we apply a circularly polarized electromagnetic wave to the plane of TI thin film, due to the anisotropy in the conductivity, the reflected and transmitted waves will not be circular anymore but they are elliptic in general. The eccentricity $e_P$ which varies from $0$ to $1$ is a measure of the stretching of the polarization ellipse and the two limiting values correspond to a circular and linear polarizations, respectively. The angle $\alpha_p$ shows the amount of the ellipse axis rotation with respect to the $x$-direction.
\par
As indicated in Fig. \ref{fig:ref} the reflected light is strongly elliptic in a wide range of magnetic fields around the transition point $\varepsilon_B=\Delta$ and the eccentricity reaches values even equal to one, corresponding to linear polarization. In addition in the region of almost linear polarization, the rotation takes place which becomes more profound for higher frequencies. Far away from the transition region $\varepsilon_B\sim\Delta$ the polarization becomes almost circular and almost untwisted since the conductivity tensor becomes isotropic and the time reversal is not broken.
\par
Similar to the reflected wave, the transmitted part becomes elliptic around $\varepsilon_B\sim\Delta$ but its eccentricity shows two peaks which their positions vary with the frequency incident wave. Nevertheless, the ellipticity of the transmitted light is limited and $e^t_P$ reaches values on the order of $\sim0.1$.
Most remarkably, the ellipse of the transmitted electromagnetic wave for small and large $B$ is aligned along the line $y=x$ and $x$-direction, indicated by rotation angles $\pi/4$ and $0$, respectively. However, for intermediate magnetic fields at the vicinity of transition point $\varepsilon_B=\Delta$, the rotation angle changes sign and becomes negative. All these features and the dependence on the frequency can be seen in Fig. \ref{fig:ref} where the eccentricity and the rotation angle are plotted as a function of the magnetic energy $\varepsilon_B$.
\par
\begin{figure}
\includegraphics[width=1.0\linewidth]{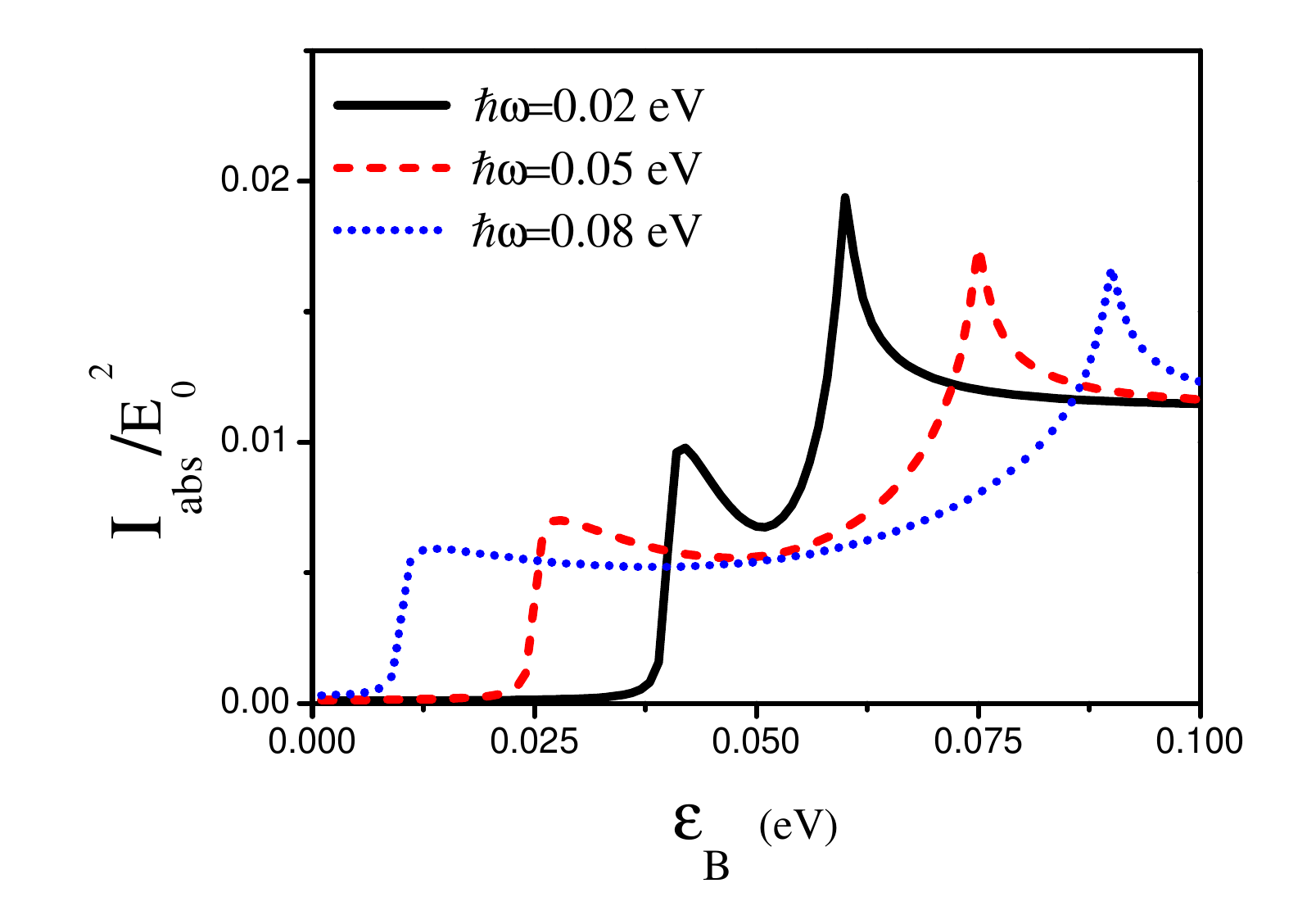}
\caption{(Color online) The relative intensity of absorbed light with respect to the incident electromagnetic wave as a function of magnetic field strength for different frequencies, $\omega$.
\label{fig:abs}}
\end{figure}
Up to now we only discussed about the optical activity of the thin film with respect to its effect on the polarization of the transmitted and reflected lights. Using Eq. \eqref{absorb} the intensity of absorbed light by the thin film can be obtained and the results are shown in Fig. \ref{fig:abs}. The ultrathin film of TI is almost transparent and does not absorb the electromagnetic waves with frequencies less than the band gap. Nevertheless by increasing the magnetic field which leads to the phase transition and entrance to the gapless phase an absorption up to a few percent is possible. Moreover, deep inside the gapless phase for very large magnetic fields ($\varepsilon_B\gg \Delta$) the absorption becomes independent of the incident light frequency. We also note that $I_{\rm abs}$ reveals a two peak structure which originates from the similar behavior of $\sigma_{yy}$ as shown in Fig. \ref{fig:q}(b),(d).

\section{Conclusions}
\label{conc}
In conclusion, we have investigated the optical responses of an ultrathin topological insulator film in the presence of an in-plane magnetic field $B$. In the absence of overlap $\Delta$ between the surface states of two sides of ultrathin film
such magnetic field can be simply gauged out and its presence will have no physical implications. However, the interplay of the overlap and in-plane magnetic field leads to the physical effect of time reversal symmetry breaking. Then the magnetic field results in the strong anisotropy of the optical conductivity especially when the magnetic energy scale and the overlap are of the same order.
At this vicinity, by increasing the magnetic field the gap of the system closes and a quantum phase transition takes place which also affects the behavior of the optical responses. For instance in the gapped phase the imaginary part of conductivities dominans while in gapless phase the dissipative real parts of $\sigma_{xx}$ and $\sigma_{yy}$ is much larger.

Our numerical results lead to the strong optical activity of the thin film TI controllable by the magnetic field strength and the thickness of the film. Assuming a normally incident circularly polarized light, the reflected and transmitted electromagnetic waves are elliptic in general. In particular, the reflected light can reach high ellipticity and even linear polarization at the vicinity of phase transition $\varepsilon_B\sim\Delta$. Therefore, the thin film can act as a polaroid for the reflected light. We should mention that since we are dealing with normally incident electromagnetic waves such result must not be mixed with well known effect described as Brewster's law. Finally it is shown that the thin film absorb the lights in the presence of the in-plane magnetic field which can reach up to a few percentage of the incident intensity.


\begin{thebibliography}{30}
\bibitem {comment} As mentioned in Ref. [\onlinecite{Sergey}] Zeeman coupling will be as important as gauge field for a thickness above $d=0.2$ nm, since our samples are thicker than $2$ nm we have neglected this term.
\end{thebibliography}
\end{document}